\documentclass[english,aps,prl,twocolumn,oneside,floatfix]{revtex4}
\usepackage[T1]{fontenc}
\usepackage[latin9]{inputenc}
\usepackage{float}
\usepackage{graphicx}
\usepackage{amssymb}

\makeatletter

\usepackage{babel}
\makeatother

\begin{document}

\title{Lifetime and Coherence of Two-Level Defects in a Josephson Junction}

\author{Yoni Shalibo$^{1}$, Ya'ara Rofe$^{1}$, David Shwa$^{1}$, Felix
Zeides$^{1}$, Matthew Neeley$^{2}$, John M. Martinis$^{2}$ and Nadav
Katz$^{1}$}

\affiliation{$^{1}$Racah Institute of Physics, The Hebrew University of Jerusalem,
Jerusalem 91904, Israel}

\affiliation{$^{2}$Department of Physics, University of California, Santa Barbara,
California 93106, USA}

\begin{abstract}
We measure the lifetime ($T_{1}$) and coherence ($T_{2}$) of two-level
defect states (TLSs) in the insulating barrier of a Josephson phase
qubit and compare to the interaction strength between the two systems.
We find for the average decay times a power law dependence on the
corresponding interaction strengths, whereas for the average coherence
times we find an optimum at intermediate coupling strengths. We explain
both the lifetime and the coherence results using the standard TLS
model, including dipole radiation by phonons and anti-correlated dependence
of the energy parameters on environmental fluctuations.
\end{abstract}


\maketitle
Two-level defects in amorphous insulators are of fundamental interest
due to their impact on many low temperature properties, such as the
heat conductivity \cite{Phillips1987} and the generation of $1/f$
noise \cite{Shnirman2005, Clare2005}. On the practical side, these
effects limit the operation of solid state devices, for example amplifiers
\cite{Kirton1989} and CCD detectors \cite{Saks1980}, and increase
the dielectric loss of insulators \cite{Martinis2005}.

Recently, a new type of solid state device has emerged in which quantum
coherence is maintained over a large distance. In particular, superconducting
Josephson qubits allow one to study quantum coherence at the macroscopic
level \cite{Shumeiko2005}. Two-level defects in their amorphous oxide
tunnel barriers (usually made of AlO$_{\mathrm{x}}$ ($\mathrm{x}\approx1$)
\cite{Tan2005}) have been found to limit the performance of these
devices \cite{Cooper2004,Martinis2005,Lucero2008}. In the phase qubit,
it was found that at certain biases the qubit is strongly coupled
to spurious two-level states (TLSs) which result in free oscillations
with the qubit and effectively reduce its coherence \cite{Simmonds2004}.
Similar effects have been observed in the flux qubit as well \cite{Lupaccu2009},
although these are less common due to its smaller junction. Supercondcuting
qubits hold promising features for the implementation of quantum information
processing devices, however a significant improvement in defect density
is required for future progress.

These defects are thought to arise from charge fluctuators in the
insulating material of the junction, presumably O-H bonds \cite{Martinis2005}.
Measurements on dielectrics at high temperature and power combined
with measurements on the phase qubit at low temperature strongly support
a two-level model for these fluctuators, emerging from tunneling between
two configurational states of the charge inside the junction. The
notion of coupling of the phase qubit to a strongly anharmonic microscopic
system was further fortified through careful analysis of the multilevel
spectrum of the phase qubit near resonance with a defect \cite{Ustinov2010}.
Neeley \emph{et al.} have demonstrated coherent control over a single
defect, and characterized its coherence and relaxation time \cite{Neeley2008b}.
Other attempts are being made to reduce the impact of defects by improving
junction materials using epitaxial growth \cite{Pappas2006} or by
fabricating dielectric free junctions \cite{Tettamanzi2006}. %
\begin{figure}[H]
\begin{centering}
\includegraphics{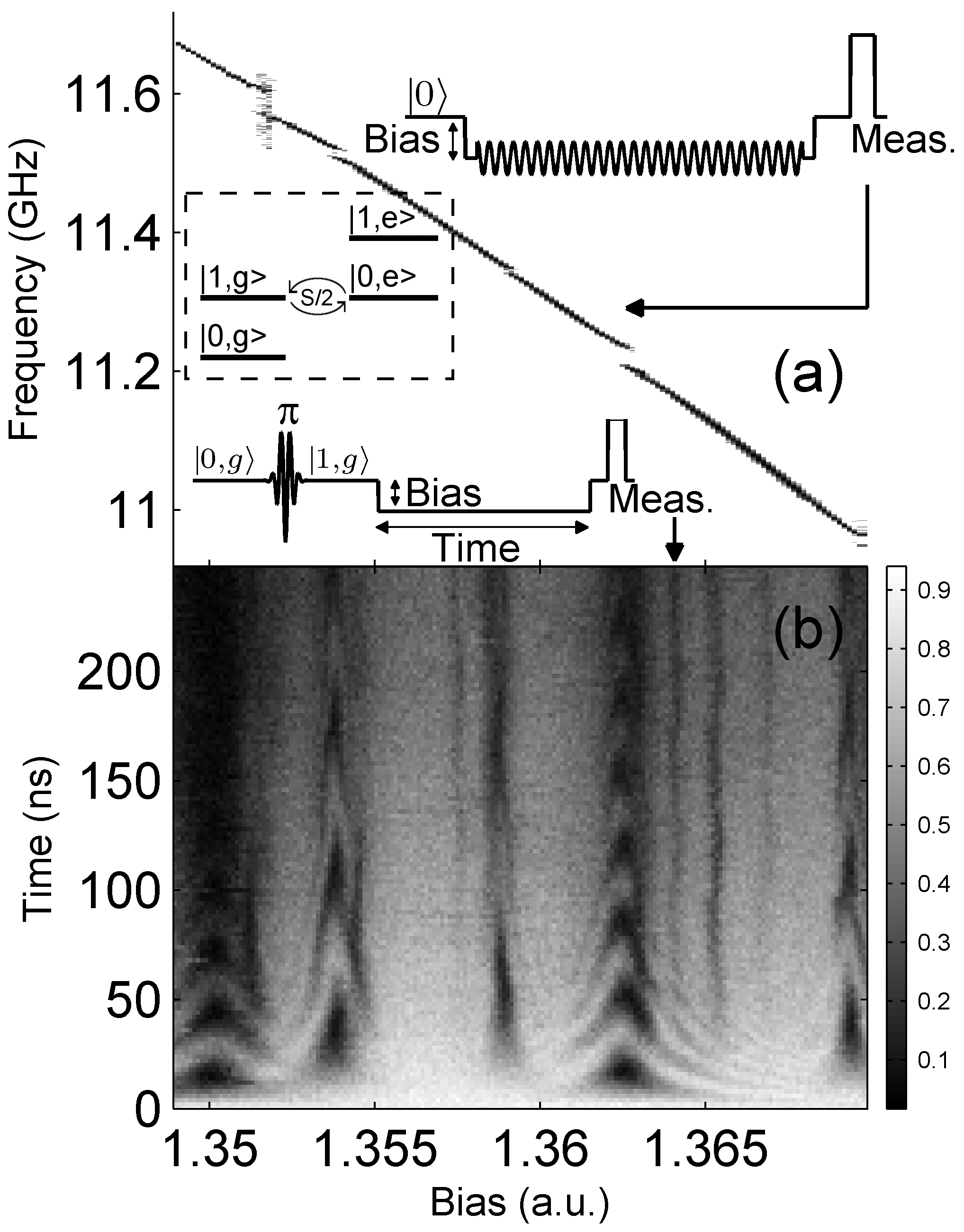}
\par\end{centering}

\caption{\label{fig:spectrum} Frequency-domain and time-domain signatures
of TLSs in qubit measurement. (a) The probability of the qubit being
excited ($P_{1}$, normalized) after a long microwave tone at different
frequencies and biases. On top of the expected smooth change of qubit
frequency with bias, we observe randomly scattered splittings due
to coupling to TLSs. (b) $P_{1}$ after qubit excitation followed
by a {}``free'' evolution at different bias values. $P_{1}$ oscillates
near resonance with TLSs, consistent with the position of splittings
in the spectrum. Upper and lower insets: control sequences used to
produce (a) and (b) respectively. $\left|0,g\right\rangle $ and $\left|1,g\right\rangle $
stand for states where the qubit is at its ground and excited state
respectively while the TLS is at its ground state. Dashed inset: level
diagram for the combined qubit-TLS system on resonance.}

\end{figure}

Several mechanisms have been proposed for relaxation and dephasing
of the dielectric defects themselves. Energy relaxation is caused
by coupling to phonon states, while dephasing could be caused by spectral
diffusion \cite{Phillips1987}. However, to date, only limited measurements
were carried out on TLSs to characterize these processes at the single
defect level \cite{Palomaki2010}. Such measurements could be used
to better understand the nature of the defects and their decay mechanisms,
and possibly engineer long-lived quantum memories in future devices.
In \cite{Palomaki2010}, the coherence times of several TLSs were
measured spectroscopically and were found to distribute as $P\sim1/T_{2}$,
where $T_{2}$ is the spectroscopic coherence time. Neeley's method
\cite{Neeley2008b} of probing the TLS adds the capability of measuring
the coherence time more accurately and also measure their lifetime
separately.

In this letter, we present a measurement of the decay of energy and
coherence for a large ensemble of TLSs in a small area junction using
the phase qubit. We find that on average, the energy relaxation time
($T_{1}$), follows a power law dependence on the coupling parameter
to the phase qubit. The exponent of this power law is in fair agreement
with what is expected from phonon radiation by a dipole (proportional
to the coupling strength) inside the junction. The average dephasing
time ($T_{\mathrm{\phi}}=\left(1/T_{2}-1/2T_{1}\right)^{-1}$) is
coupling dependent as well, peaking at intermediate couplings. We
interpret this optimum coupling to be caused by anti-correlated fluctuations
in the physical parameters which determine the TLS energy.

For small area junctions ($\sim1\,\mu m^{2}$), the typical measurement
bandwidth allows us to detect and measure about 10 TLSs in a particular
cooldown. Instead of using many different samples to acquire sufficient
statistics, we use the fact that heating resets the TLS characteristics.
The device is thermally anchored to the mixing chamber of a dilution
refrigerator during measurement. We find that the TLS distribution
is reset upon raising the temperature above 20\,K and cooling down
to the base temperature (10\,mK). Some memory of the TLS distribution
remains if the temperature is increased to only 1.5\,K \cite{SuppMat}.
We utilize this feature to produce a new set of TLSs and generate
an ensemble. The data was taken over 82 different TLSs, obtained from
8 different cooldowns.

The initial identification of TLSs and their coupling parameters are
carried out as follows. First, the qubit spectrum is swept over the
bias %
\footnote{The qubit loop is coupled to an external bias source through a flux
transformer, which sets the qubit energy \cite{Martinis2002}.%
} to locate the frequencies of TLSs from the positions of the avoided
level crossing structures (see Fig. \ref{fig:spectrum}a). A complementary
picture of the interacting qubit-TLS system in the time domain is
shown in Fig. \ref{fig:spectrum}b, where we excite the qubit with
a short resonant pulse ($\pi$-pulse) far away from any observable
TLS and then apply a bias pulse of varying amplitude and length. As
seen in the figure, for bias values where the qubit is resonant with
a TLS we observe oscillations that have the same frequency as the
splitting size in the spectrum. %
\begin{figure}
\begin{centering}
\includegraphics{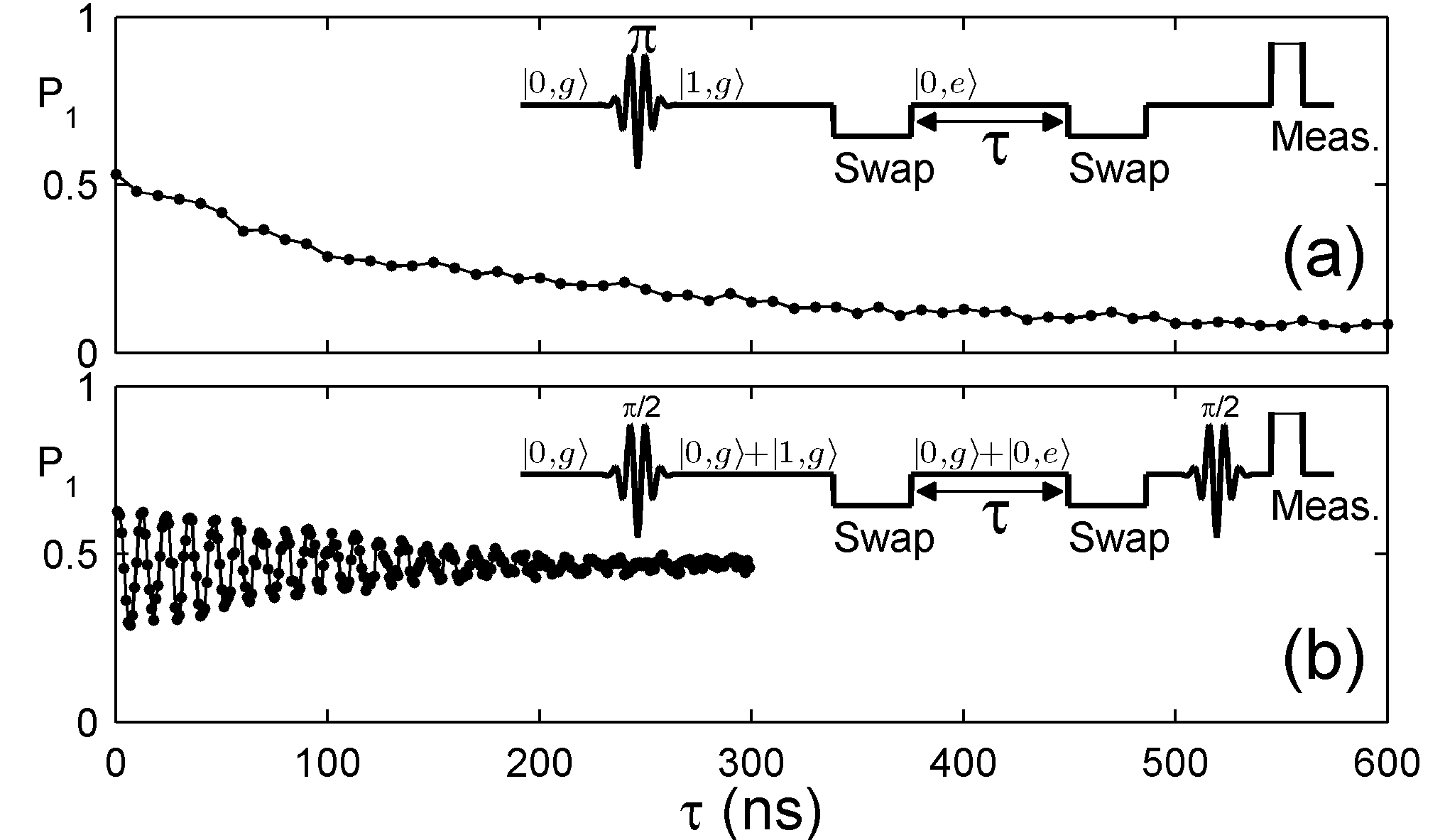}
\par\end{centering}

\caption{\label{fig:decays} Representative $T_{1}$ and $T_{2}$ measurements
of a TLS. (a) $T_{1}$ measurement of the TLS, along with its experimental
sequence (inset). The qubit is first excited with a $\pi$-pulse,
then brought into resonance with a TLS for a {}``swap time'' (the
time to fully transfer an excitation between the qubit and the TLS.
It is found for each TLS by locating the first minimum in the oscillations
in Fig. \ref{fig:spectrum}b). After a free evolution of the TLS of
time $\tau$, a swap gate is again applied, afterwhich the qubit excitation
probability $P_{1}$ is measured. (b) $T_{2}$ measurement of a TLS,
along with its experimental sequence (inset). This sequence is similar
to a $T_{1}$ measurement, only that superposition states are produced
in the TLS and their decay is measured vs. time (Ramsey sequence \citep{Neeley2008b}).
The amplitude of the oscillations is proportional to the degree of
coherence in the TLS.}

\end{figure}

\begin{figure}
\begin{centering}
\includegraphics{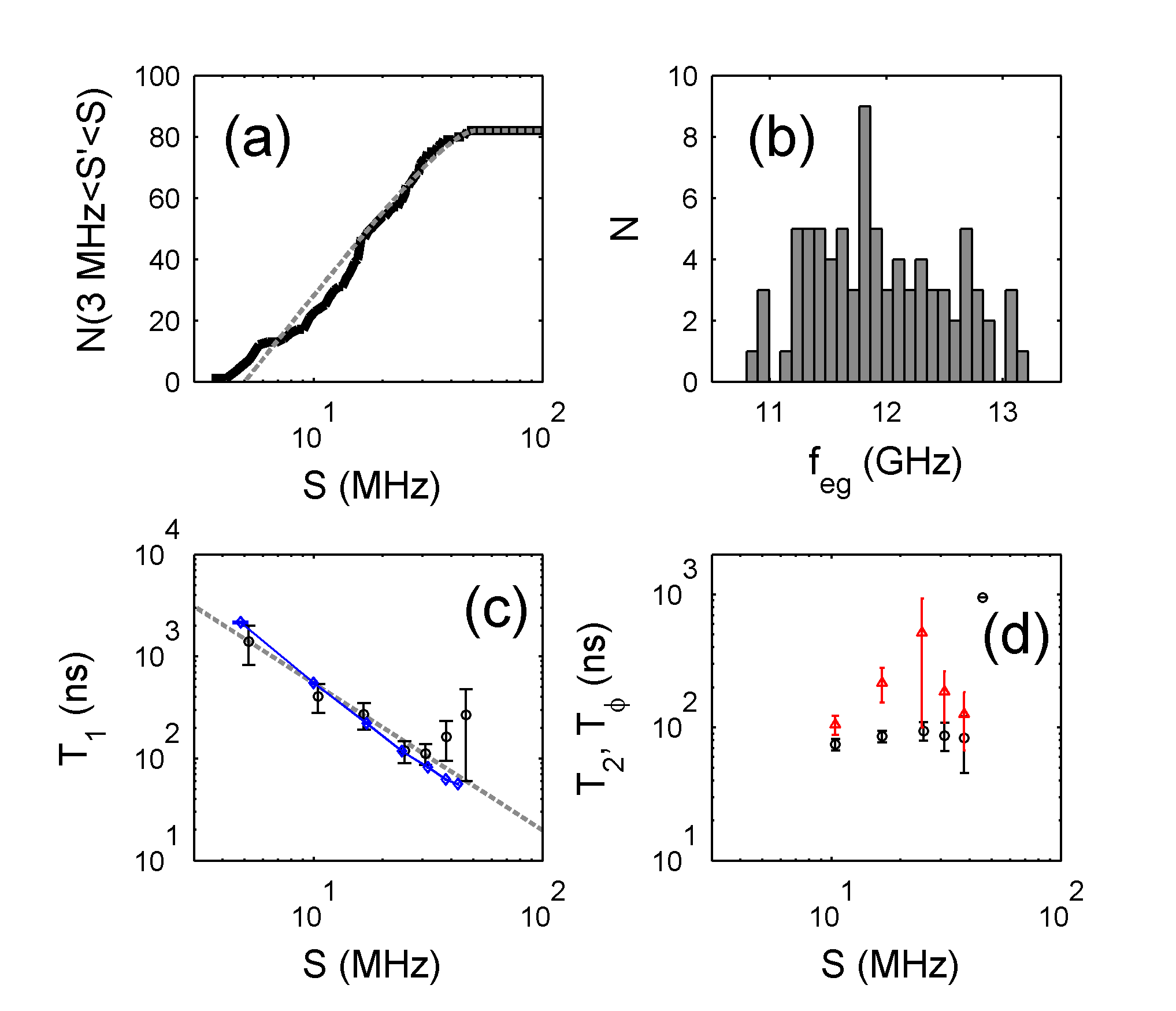}\caption{\label{fig:survey} TLS survey results. (a) Frequency and (b) splitting-size
distribution of 82 TLSs. The curve in (a) is the best-fitted log-normal
distribution of the standard TLS model \citep{Martinis2005}. (c)
Average $T_{1}$ values (black circles) as a function of average splittings
taken for TLSs inside a 7\,MHz splitting window, and the best fitted
power law in dashed gray. Average $T_{1}$ from stochastic simulation
(arbitrary amplitude is shown) in blue diamonds. We attribute the
deviation at the largest splittings to low statistics within these
windows \citep{SuppMat} (d) Average $T_{2}$ (black circles) and $T_{\phi}$
(red triangles) value as a function of average splitting size within
a 7\,MHz window.}

\par\end{centering}
\end{figure}

Following Neeley \emph{et al.} \cite{Neeley2008b}, the characteristic
energy relaxation and decoherence time scales were extracted from
$T_{1}$ and Ramsey experiments on the TLS, with sequences schematically
represented in the insets of Fig. \ref{fig:decays}. Figure \ref{fig:decays}a
and \ref{fig:decays}b show representative $T_{1}$ and $T_{2}$ decay
curves of the same TLS with characteristic times of 187\,ns and 148\,ns
respectively, obtained from a fit to a decaying exponent and an oscillatory
decaying exponent. The size distribution of the observed splittings
(see Fig. \ref{fig:survey}a) follows the theoretical curve predicted
by the standard model for two-level defects and agrees with previous
results on similar junctions (generated by measuring different samples)
\cite{Martinis2005}. The maximal splitting size is found to be 45\,MHz.
Theory predicts \cite{Martinis2005} that this maximal splitting $S_{\mathrm{max}}$
depends on junction parameters and defect size according to $S_{\mathrm{max}}=\left(2d/x\right)\sqrt{e^{2}E_{10}/2C}$
, where $x$ is the barrier thickness, $d$ is the spatial size of
the dipole, $C$ is the junction capacitance and $E_{10}$ is the
qubit energy. From the measured $S_{\max}$ and known junction parameters
we compute a dipole size $d\simeq1\,\textrm{\AA}$. The minimal observable
splitting size is $\sim3$\,MHz, and is mainly limited by the coherence
time of the qubit. In addition, we find the distribution of TLS energies
($E_{\mathrm{ge}}$, the energy between the ground state and excited
state) to be constant throughout our qubit measurement bandwidth (see
Fig. \ref{fig:survey}b), consistent with theory.

Although most of the $T_{1}$ decay curves of the TLSs are similar
in their shape (i.e. a simple exponential decay), their decay times
range almost 3 orders of magnitude, from 12\,ns to more than 6000\,ns.
Coherence times on the other hand range from 30\,ns to only 150\,ns
(excluding a single anomalous TLS which will be discussed later).
For comparison, when the qubit is biased far from any observable splitting,
its lifetime is 270\,ns, and its coherence time is 90\,ns. TLS energy
relaxation times at a given splitting are not random. We find that
they are shorter for larger splittings (stronger interaction with
the qubit), although short lifetimes are measured for the smallest
splittings as well. This trend is apparent in Fig. \ref{fig:survey}c
where we plot average $T_{1}$ values as a function of splitting.
In this plot we divide the ensemble into groups of TLSs having similar
splitting values, in a 7\,MHz window size. The error bar represents
the statistical spread of the data within this window \cite{SuppMat}.
We find the average values $\left\langle T_{1}(S)\right\rangle $,
excluding two points \cite{SuppMat}, to be best fitted by a power
law $T_{1}\varpropto S^{\alpha}$, where $\alpha=-1.44\pm0.15$ \cite{SuppMat}.
Figure \ref{fig:survey}d (black circles) shows the processed $T_{2}$
data, obtained similarly from only 43 different TLSs \cite{SuppMat}.
In this case we observe a weak dependence on the coupling with a peak
at $S\approx25$\,MHz. This feature is more pronounced in the dephasing
time $T_{\mathrm{\phi}}$, represented by red triangles in Fig. \ref{fig:survey}d.

The $T_{1}$ results can be understood within the standard TLS model.
The excited state of the TLS involves a local deformation of the insulator.
This deformation couples to phonon modes, leading to the decay of
the TLS excitation. The expected lifetime for such a process \cite{Phillips1987},
is given by

\begin{equation}
T_{1}^{-1}=\frac{E_{\mathrm{ge}}\Delta_{0}^{2}\gamma^{2}}{2\pi\rho\hbar^{4}}\left(\frac{1}{v_{\mathrm{l}}^{5}}+\frac{2}{v_{\mathrm{t}}^{5}}\right),\label{eq:T1}\end{equation}
where $\gamma$ is the deformation potential, $v_{\mathrm{l}}$ and
$v_{\mathrm{t}}$ are the speeds of sound for the longitudinal and
transverse modes respectively, $\Delta_{0}$ is the energy splitting
due to tunneling and $\rho$ is the mass density. This is consistent
with a power law dependence on $S$, since the interaction strength
with the qubit satisfies $S\propto S_{\mathrm{max}}\Delta_{0}/E_{\mathrm{ge}}$
\cite{Martinis2005}.

The interaction of a TLS with the qubit is that of an electric dipole
with an electric field, and therefore depends on the dipole orientation
\cite{Martinis2005}. This feature explains the large spread in the
data at a given splitting: both large dipoles (large $\Delta_{0}/E_{\mathrm{ge}}$)
perpendicular to the junction's electric field and small dipoles aligned
with the field can have the same $S$ but different lifetimes. To
more rigorously compare experiment with this theory, we simulate an
ensemble of TLSs with uniform distribution of dipole orientation and
log distribution of dipole moment sizes \cite{Martinis2005}, from
which we calculate the average lifetimes as a function of splitting
size \cite{SuppMat}. The simulation data (see Fig. \ref{fig:survey}c)
yields an average exponent $\alpha=-1.63$ \cite{SuppMat}, in agreement
with our measurement.  %

The magnitude of the times that we extract from the experiment at
a given splitting can be compared to the expected values for defects
inside an AlO$_{\mathrm{x}}$ dielectric using Eq. \ref{eq:T1}. We
approximate the deformation potential by $\gamma=\frac{1}{2}\rho v^{2}\Delta V$
\cite{Sethna1982}, where $v$ is the average speed of sound of the
transverse and longitudinal modes and $\Delta V$ is the local difference
in volumes. For Al$_{2}$O$_{3}$ values with the difference in volumes
taken as $\Delta V\simeq a^{3}$, where $a\simeq1\,\textrm{\AA}$
is the extracted dipole size, we get $\gamma\approx$1\,eV, consistent
with defects in other dielectrics \cite{Clare2005}. Since the dielectric
layer thickness is much smaller than the relevant phonon wavelength,
the speed of sound in Eq. \ref{eq:T1} is set by the aluminum layers
of the electrodes. Using the speed of sound for thin aluminum films
\cite{CRC}, we get $T_{1}(S_{\mathrm{max}})\simeq30$\,ns which
is very similar to what we measure. A more specific estimation should
take into account the size of the junction and the layered structure.

Assuming the dephasing process is caused by fluctuations in energy,
we note that the maximum observed in Fig. \ref{fig:survey}d can be
explained by an anti-correlated dependence of the charging energy
and tunneling energy on fluctuations in the TLS environment. According
to the TLS model, $E_{\mathrm{ge}}=\sqrt{\Delta^{2}+\Delta_{0}^{2}}$
where $\Delta$ is the energy difference between the bare states of
two spatial configurations $\left|L\right\rangle $ and $\left|R\right\rangle $
and $\Delta_{0}$ is the tunneling interaction energy. Both $\Delta$
and $\Delta_{0}$ are dependent on a set of environmental parameters
$\overrightarrow{P}$, which fluctuate in time. As is standard for
the TLS model, we assume a linear sensitivity for $\Delta$ on $\overrightarrow{P}$
and an exponential sensitivity for $\Delta_{0}$: $\Delta_{0}(\overrightarrow{P})=N_{1}e^{-\sum P_{i}/P_{0i}},\Delta(\overrightarrow{P})=N_{2}\sum P_{i}/P_{1i},$
with overall dimensional normalization constants $N_{1}$ and $N_{2}$
and parameter specific constants $\overrightarrow{P_{0}}$ and $\overrightarrow{P_{1}}$.
The resulting fluctuations in energy, to first order in fluctuations
$\delta\overrightarrow{P}$ in these parameters, are given by $\delta E_{\mathrm{ge}}=\frac{1}{E_{\mathrm{ge}}}\sum\delta P_{i}\left(\Delta^{2}/P_{1i}-\Delta_{0}^{2}/P_{0i}\right)$.
This expansion becomes interesting for the situation where $\Delta_{0}\sim\Delta$
as there is a possibility for the contribution in $\delta E_{\mathrm{ge}}$
to pairwise cancel. Since $\Delta_{0}=S\left(E_{\mathrm{ge}}/S_{\mathrm{max}}\cos\eta\right)$,
where $\eta$ is the dipole orientation relative to the junction's
electric field, we expect to find such a cancellation at a particular
splitting $S$. Note that the dependence on $\cos\eta$ smears this
somewhat but we still expect a significant effect, as is observed
in Fig. \ref{fig:survey}d.

As seen in the figures, the power-law describing $T_{1}(S)$ cannot
explain all the measured TLSs. We find that three out of 82 TLSs with
large splittings (37\,MHz, 41\,MHz and 45\,MHz) have much longer
lifetimes than expected (220\,ns, 243\,ns and 476\,ns - respectively).
In addition, one TLS out of 41 has much longer coherence than all
the others (about a factor of 6 longer than the longest $T_{2}$ of
all the others), associated with a splitting of size 30\,MHz. Other
anomalies we discovered are related to the stability of a particular
TLS in time. We find that the energy $E_{\mathrm{ge}}$ of some TLSs
(about 5\,\%) changes spontaneously at varying time scales, from
seconds to days. All the rest were remarkably stable \cite{SuppMat}.

Some of these changing TLSs have long lifetime (a few microseconds),
which is consistent with the power-law trend we discussed above. Furthermore,
we also measure a few representative TLSs as a function of temperature.
We find no significant change in $T_{1}$ and $T_{2}$ below 100\,mK,
consistent with the expected \cite{Phillips1987} $\tanh(E_{\mathrm{ge}}/2k_{\mathrm{B}}T)$
dependence. We also find that the instability of some TLSs increases
at elevated temperatues (i.e, the change in TLS energy becomes more
frequent). We conclude that some of the TLSs we measure have a different
nature, perhaps related to their internal structure or position inside
the junction.

In conclusion, the energy decay and dephasing times of two-level defects
in an AlO$_{\mathrm{x}}$ barrier of a Josephson junction are measured
as a function of the coupling parameter with the phase qubit. The
lifetimes vary substantially in our range of splittings, and agree
with the theoretically predicted phonon radiative loss, which is dipole
size dependent. The dephasing times show an extremum at intermediate
couplings, which we attribute to an anti-correlated dependence on
fluctuations in the environmental parameters which set the TLS energy.
Such a dependence may distinguish between different theoretical models
for TLSs. Our results demonstrate the power of the phase qubit as
a dynamical coupling element to microscopic systems at the single
microwave photon level.

We acknowledge fruitful discussions with Clare C. Yu. This work was
supported by ISF grant 1835/07 and BSF grant 2008438.

\end{document}

\title{Supplementary Material}

\maketitle

\section{Experimental data}

\begin{figure}[H]
\includegraphics{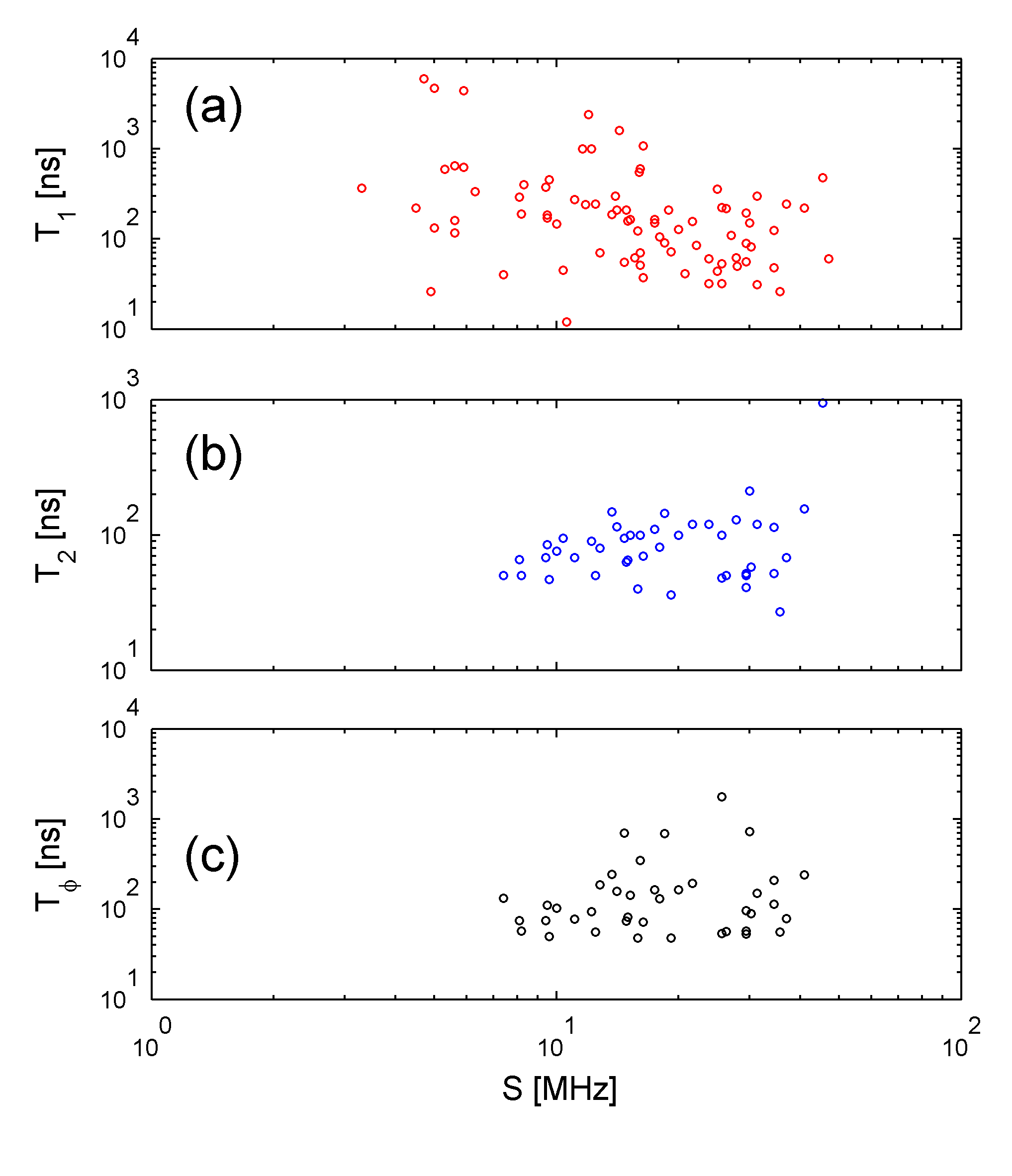}\caption{\label{fig:FullSurvey}(a) Measured $T_{1}$ \emph{vs.} splitting,
(b) $T_{2}$ \emph{vs.} splitting and (c) $T_{\phi}$ \emph{vs.} splitting
for all the measured TLSs. The spread in the $T_{1}$ data at a particular
splitting value results from the random distribution of TLS orientation
in the junction. It is apparent that the maximal lifetime shortens
at larger splittings, consistent with dipole radiation. The spread
of the $T_{2}$ data appears independent of the splitting, however
the average values show some dependence which becomes more prominent
in the dephasing times. }

\end{figure}

The full data measured for 82 TLSs is presented in Table \ref{tab:data}
and the $T_{1}(S)$, $T_{2}(S)$ and $T_{\phi}(S)$ values are plotted
in figure \ref{fig:FullSurvey}. Figure \ref{fig:FullSurvey}a contains
82 data points while \ref{fig:FullSurvey}b contains only 42. For
about 50\,\% of the cases the dephasing time could not be determined,
due to both low visibility and short dephasing time. This happens
mostly for small splittings, which do not lie in the range of points
presented in Fig. 3d of the main paper. For points that lie within
this range and are omitted from the analysis we separately checked
that the shorter coherence time does not affect the trend. The number
of points in Fig. \ref{fig:FullSurvey}c is further reduced because
for some $T_{1}$ limited TLSs we measured $T_{2}$ which is slightly
longer than $2T_{1}$ due to measurement error. $T_{\phi}$ is excluded
from the figure and table for these cases (three TLSs).%

The errors in Fig. 3b and and Fig. 3c in the main paper represent
the statistical spread of the data within a 7\,MHz window. They are
calculated by normalizing the standard deviation by $\sqrt{N}$, where
$N$ is the number of points within the window. %
\begin{figure}[H]
\includegraphics{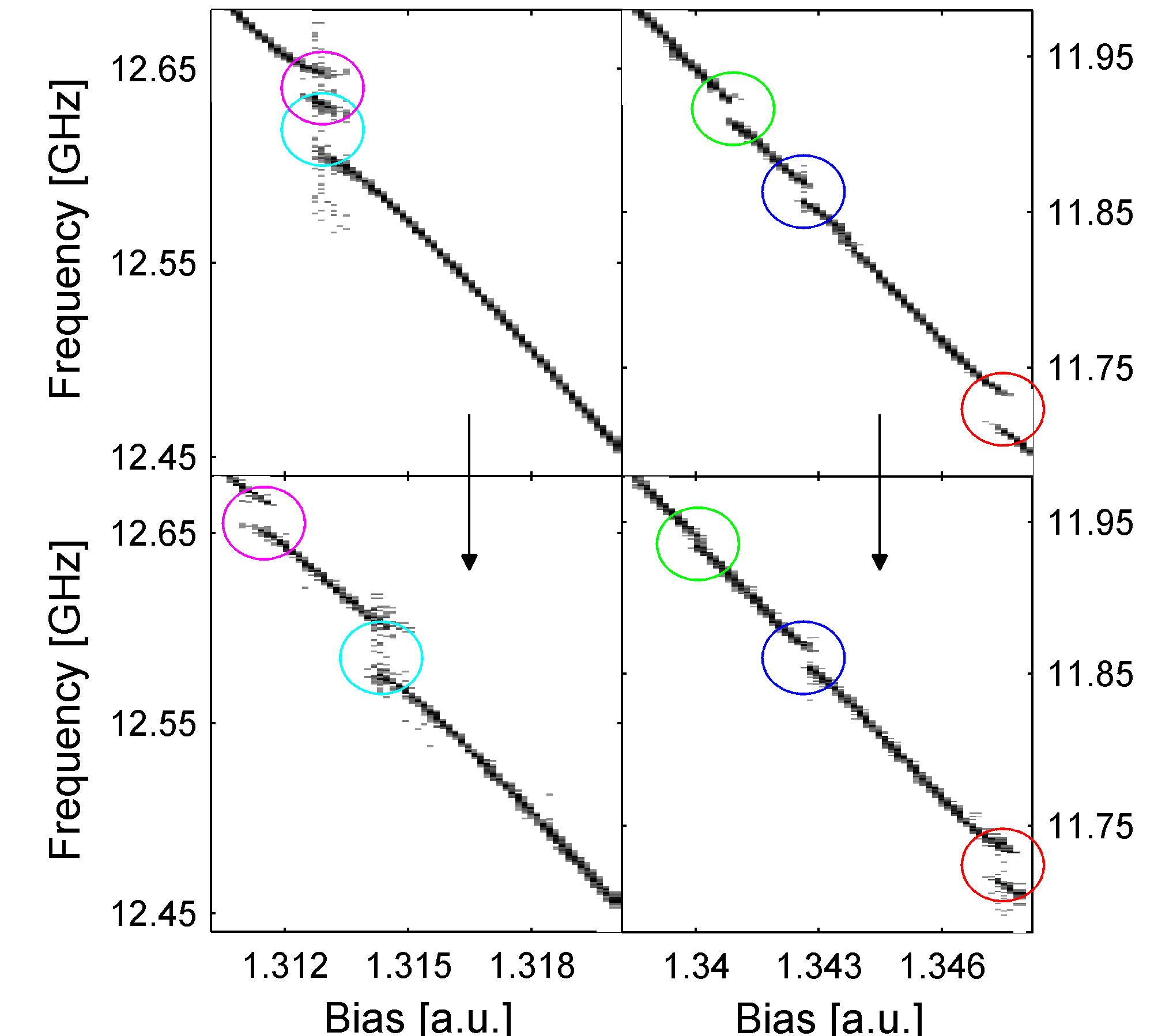}\caption{\label{fig:PartialWarmup} Qubit spectrum as a function of bias, before
and after warmup to 1.5\,K. Upper: two sections of the spectrum before
warmup. Lower: the same sections, taken after heating the sample to
1.5\,K and cooling down back to 10\,mK. Circles with the same color
indicate a splitting that we attribute to the same TLS. Both the position
in frequency and splitting size of the TLSs are similar, indicating
that the TLSs are not fully reset at this temperature.}

\end{figure}

We would like to point out that two points in Fig. 3c in the main
paper have been excluded from the fit to a power law. These are the
points with the largest average splittings (38\,MHz and 46\,MHz),
where the statistics within each window is low (3 data points and
2 data points respectively). If we include these points in the fit
we obtain an exponent $\alpha=-0.93$. We believe that the data points
at the largest splittings may result from anomalous TLSs.

As pointed out in the main paper, some of the TLSs are changing in
time. This is characterized by a change in the TLSs' transition energy,
causing them to disappear from time to time (that is, to step out
of our measurement bandwidth) or appear at a slightly different energy.
An example of this phenomenon can be seen in Fig. 1 of the main paper.
The leftmost TLS appears in the spectrum at larger bias value (that
is smaller transition energy) than in the time-domain sweep. These
two measurements were taken at an interval of one day. 

New sets of TLSs are produced from the same device by warming up to
20\,K. We believe that some TLSs are not fully reset after warming
up to only 1.5\,K. Some splittings in the spectrum are similar in
their frequency and splitting size to those before a partial warmup,
as indicated in Fig. \ref{fig:PartialWarmup}. Both the reset of TLSs
at high temperatures, and the fact that some of the TLSs are changing
in time indicate the true nature of the TLS as an approximation of
a multilevel state, resulting from a multi-well energy structure.

\section{Stochastic simulation}

We reproduced the lifetime distribution of an ensemble of TLSs according
to the TLS model in the following way. For each TLS we assume a uniform
distribution of dipole orientation ($\cos\eta$ distributes uniformly,
where $\eta$ is angle relative to the electric field inside the junction),
a potential asymmetry $\Delta$ from a uniform distribution ($\Delta\varpropto z$,
where $z$ is the effective distance between position states of the
TLS inside the junction) and a tunneling energy $\Delta_{0}$ from
a log distribution ($\ln\Delta_{0}\varpropto z$). These distributions
are consistent with the TLS model, which assumes linear sensitivity
of $\Delta$ on $z$ and exponential sensitivity of $\Delta_{0}$
on $z$. We excluded TLSs having smaller couplings than we can measure.
For each TLS we compute the lifetime according to Eq. 1 in the main
paper ($T_{1}(\sin(\theta))=a/\sin(\theta)^{2}$, where $\theta=\arctan(\Delta_{0}/\Delta)$
and $a$ is some constant). The simulation data points were then averaged
over a 7\,MHz window size, as done for the experimental data.

The resulting $\left\langle T_{1}(S)\right\rangle $ behavior (see
Fig. 3c in the main paper, blue diamonds) resembles a sum of two power
laws. At smaller splittings $S\lesssim S_{max}/2$, the points fit
a power law with an exponent $\alpha_{1}\approx-1.9$, while for larger
splittings they fit a power law with an exponent $\alpha_{2}\approx-1$.
A similar trend is observed in our data as well. 

\begin{table*}
\begin{tabular}{r|r|r|r|r|r}
$f_{\mathrm{ge}}$ [GHz] &   $S$ [MHz] &        $T_{1}$ [ns] &          $T_{2}$ [ns]&          $T_{\phi}$ [ns] &          CD No. \\
\hline 
      12.8 &        3.3 &        365 &          - &          - &          6 \\
     11.43 &        4.5 &        220 &          - &          - &          3 \\
     11.35 &        4.7 &       6000 &          - &          - &          7 \\
     12.45 &        4.9 &         26 &          - &          - &          7 \\
     11.55 &        5.0 &       4700 &          - &          - &          4 \\
     11.31 &        5.0 &        132 &          - &          - &          7 \\
     11.23 &        5.3 &        590 &          - &          - &          4 \\
    11.462 &        5.6 &        650 &          - &          - &          3 \\
     13.08 &        5.6 &        117 &          - &          - &          4 \\
     11.33 &        5.6 &        160 &          - &          - &          4 \\
     12.89 &        5.9 &       4400 &          - &          - &          6 \\
      11.4 &        5.9 &        623 &          - &          - &          6 \\
     12.79 &        6.3 &        335 &          - &          - &          4 \\
     13.11 &        7.4 &         40 &         50 &        133 &          4 \\
     12.85 &        8.1 &        291 &         66 &         74 &          3 \\
     12.13 &        8.2 &        190 &         50 &         58 &          5 \\
     11.57 &        8.3 &        400 &          - &          - &          4 \\
     11.45 &        9.4 &        373 &         68 &         75 &          5 \\
    11.835 &        9.5 &        170 &          - &          - &          2 \\
    11.342 &        9.5 &        185 &         85 &        110 &          2 \\
     12.26 &        9.6 &        453 &         47 &         50 &          2 \\
     12.32 &       10.0 &        147 &         76 &        102 &          5 \\
     11.85 &       10.4 &         45 &         95 &          - &          4 \\
      11.6 &       10.6 &         12 &          - &          - &          6 \\
     12.25 &       11.1 &        275 &         68 &         78 &          7 \\
        12 &       11.6 &       1000 &          - &          - &          4 \\
     11.86 &       11.8 &        240 &          - &          - &          2 \\
     11.22 &       12.0 &       2400 &          - &          - &          1 \\
     11.88 &       12.2 &       1000 &         90 &         94 &          5 \\
     12.57 &       12.5 &        243 &         50 &         56 &          6 \\
     11.96 &       12.8 &         70 &         80 &        187 &          3 \\
    11.515 &       13.7 &        187 &        148 &        245 &          2 \\
     11.96 &       14.0 &        300 &          - &          - &          4 \\
     10.96 &       14.1 &        209 &        115 &        159 &          2 \\
      10.8 &       14.3 &       1600 &          - &          - &          7 \\
     13.22 &       14.7 &         55 &         95 &        697 &          6 \\
     11.57 &       14.9 &        210 &         63 &         74 &          3 \\
     11.27 &       15.0 &        158 &         65 &         82 &          5 \\
     12.24 &       15.2 &        165 &        100 &        143 &          3 \\
      11.7 &       15.6 &         62 &          - &          - &          4 \\
     11.38 &       15.9 &        123 &         40 &         48 &          7 \\
     11.59 &       16.0 &        550 &          - &          - &          7 \\
     11.78 &       16.1 &         51 &          - &          - &          5 \\
     11.62 &       16.1 &         70 &        100 &        350 &          5 \\
     10.95 &       16.1 &        600 &          - &          - &          1 \\
\end{tabular}   \begin{tabular}{r|r|r|r|r|r}
$f_{\mathrm{ge}}$ [GHz] &   $S$ [MHz] &        $T_{1}$ [ns] &          $T_{2}$ [ns] &          $T_{\phi}$ [ns] &          CD No. \\
\hline 
    12.642 &       16.4 &       1080 &         70 &         72 &          2 \\
     11.85 &       16.4 &         37 &          - &          - &          7 \\
     12.41 &       17.5 &        165 &        110 &        165 &          5 \\
     12.37 &       17.5 &        150 &          - &          - &          7 \\
    12.542 &       18.0 &        106 &         81 &        131 &          8 \\
     11.42 &       18.5 &         91 &        144 &        690 &          4 \\
    11.915 &       18.9 &        208 &          - &          - &          2 \\
     12.47 &       19.2 &         72 &         36 &         48 &          6 \\
     11.21 &       20.0 &        127 &        100 &        165 &          5 \\
     12.73 &       20.8 &         41 &          - &          - &          7 \\
     11.86 &       21.7 &        156 &        120 &        195 &          2 \\
     12.15 &       22.2 &         85 &          - &          - &          7 \\
     11.72 &       23.8 &         60 &        120 &          - &          2 \\
     12.66 &       23.8 &         32 &          - &          - &          7 \\
     12.44 &       25.0 &         44 &          - &          - &          6 \\
     11.86 &       25.0 &        355 &          - &          - &          1 \\
     11.61 &       25.6 &         53 &        100 &       1767 &          3 \\
     12.34 &       25.6 &         32 &          - &          - &          6 \\
     11.73 &       25.6 &        224 &         48 &         54 &          7 \\
     13.04 &       26.3 &        217 &         50 &         57 &          6 \\
    12.675 &       27.0 &        110 &          - &          - &          1 \\
     11.77 &       27.8 &         62 &        130 &       2687 &          4 \\
      12.2 &       28.0 &         50 &          - &          - &          1 \\
    12.613 &       29.4 &         56 &         52 &         97 &          2 \\
     11.22 &       29.4 &         89 &         41 &         53 &          2 \\
     12.67 &       29.4 &        193 &         50 &         57 &          3 \\
    12.147 &       30.0 &        150 &        212 &        723 &          8 \\
     12.34 &       30.3 &         82 &         58 &         90 &          4 \\
     11.89 &       31.3 &         31 &          - &          - &          6 \\
     10.91 &       31.3 &        300 &        120 &        150 &          7 \\
     12.78 &       34.5 &        125 &        114 &        210 &          5 \\
     12.12 &       34.5 &         48 &         52 &        113 &          6 \\
     11.59 &       35.7 &         26 &         27 &         56 &          2 \\
     11.17 &       37.0 &        243 &         68 &         79 &          6 \\
    11.772 &       41.0 &        220 &        156 &        242 &          1 \\
     12.05 &       45.5 &        476 &        950 &          - &          4 \\
     11.99 &       47.0 &         60 &          - &          - &          1 \\
\end{tabular}   \caption{\label{tab:data}Full measurement data of TLSs: TLS energy ($f_{\mathrm{ge}}$),
splitting ($S$), lifetime ($T_{1}$), coherence time ($T_{2}$) calculated
dephasing time ($T_{\phi}$) and cooldown number. The last column
indicates which TLSs were measured on the same cooldown (i.e., temperature
was not raised to more than 20\,mK between measurements of TLSs belonging
to the same cooldown).}

\end{table*}